# Mitosis, Cytoskeleton Regulation, and Drug Resistance in Receptor Triple Negative Breast Cancer

Alexandre Matov[1, †]

1 DataSet Analysis LLC, 155 Jackson St, San Francisco, CA 94111

† Corresponding author:

email: matov@datasetanalysis.com

**ABSTRACT**

**Introduction:** Methods for personalizing medical treatment are the focus of contemporary biomedical research. In cancer care, we can analyze the effects of therapies at the level of individual cells. Quantitative characterization of treatment efficacy and evaluation of why some individuals respond to specific regimens, whereas others do not, require additional approaches beyond genetic sequencing at single time points. Methods for the continuous analysis of changes in phenotype, such as *in vivo* and *ex vivo* morphology and motion tracking of cellular proteins and organelles, over time frames spanning minute-to-hour scales, can provide important insights into patient treatment options.

**Methods:** Despite improvements in the diagnosis and therapy of many types of breast cancer (BC), many aggressive forms, such as receptor triple-negative cancers, are associated with the worst patient outcomes. Although initially effective in reducing tumor burden for some patients, acquired resistance to cytotoxic chemotherapy is almost universal, and there is no rationale for identifying intrinsically drug-resistant and drug-sensitive patient populations before initiating therapy. During cell division, the receptor triple-negative MDA-MB-231 mitotic spindles are the largest compared to other BC cell lines. Many of the MDA-MB-231 spindles exhibit rapid lateral twisting during metaphase, which remains unaffected by knockdown of the oncogene Myc and treatment with inhibitors of the serine/threonine-protein kinase B-Raf and the epidermal growth factor receptor, either alone or in any combination.

**Results:** MDA-MB-231 cells are the most aggressive and rapidly form metastatic tumors in xenograft transplant models. They exhibited very high proliferation rates when plated as three-dimensional cultures in Matrigel. Quantitative image analysis of microtubules (MTs) in six BC cell lines - MDA-MB-231 (receptor negative), HCC-1143 (receptor negative), HCC-3153 (receptor negative), ZR-75-B (estrogen receptor-positive), LY-2 (progesterone receptor-positive), HCC-1428 (estrogen receptor-positive, progesterone receptor-positive) - demonstrated that the rotational spindle rocking of MDA-MB-231 cells during metaphase appears coupled with a significant increase in MT polymerization rates during interphase, which likely shortens interphase and accelerates cell cycle progression and mitotic entry. Unlike the uniform treadmilling rates of about 21 μm/min in kinetochore MTs during metaphase we measured across cell lines, MDA-MB-231 cells in interphase exhibit the fastest MT polymerization dynamics of about 19 μm/min and this is coupled with abnormal mitotic spindle oscillations of almost 30 μm/min.

**Conclusions:** The aberrant behavior in MDA-MB-231 spindles may represent a therapeutically targetable disrupted mechanism of spindle positioning in receptor triple-negative BC (TNBC) cells leading to tumor aggressiveness. In this manuscript, we outline a strategy for selecting the most optimal tubulin inhibitor based on its ability to affect MT dynamics.

## INTRODUCTION

When epithelial cells become neoplastic, this is reflected in changes in their ability to remodel their cytoskeleton and respond to drug treatment. The cellular highways, which stretch from the nucleus to the cell cortex, are the microtubules (MTs). MTs are essential polymer filaments composed of tubulin subunits that organize and rearrange the interior of the cell (Inoué, 1959; Inoué and Salmon, 1995; Lyle et al., 2009a; Lyle et al., 2009b). They play indispensable roles in several cellular processes such as migration, mitosis, and internal transport, which are critically involved in cancer metastasis and proliferation. MTs are structurally stiff but dynamic. Their tendency to rapidly extend and retract in a stochastic way is termed dynamic instability of MT plus tips or plus ends (Mitchison and Kirschner, 1984). This dynamicity allows a cell to utilize MTs in diverse functional contexts, one of which is mitosis. Within the spindles, MTs undergo a process termed treadmilling (Maddox et al., 2003) during which the plus ends are located in proximity to the chromosomes at the metaphase plate, where MTs add dimers or polymerize, and the minus ends are located outward toward the two poles, from where the MTs remove dimers or depolymerize at similar rates. These MT-based structures align duplicated chromosomes near the cell center and ultimately segregate the copies into each of the daughter cells. In interphase, MTs are the freeways of the cellular transport infrastructure. Failure to maintain the appropriate MT network in normal physiology is a hallmark of cancer (Hanahan and Weinberg, 2000; Hanahan and Weinberg, 2011) and a number of other diseases. Tubulin inhibitors bind to MTs and induce the MTs to modify their intercellular organization. Changing the arrangement of MTs alters how and where cellular signals are delivered which, in turn, changes MT and cellular homeostasis.

Our objective has been to elucidate the mechanisms of regulation by which cancer cells affect MT organization and how they affect the susceptibility of the cell to tubulin inhibitors (Cobham and Donovan, 2009; Cortes et al., 2011; Joensuu et al., 2006; Martin et al., 2005; Sparano et al., 2008; Urruticoechea et al., 2005). To achieve this goal, we have developed software tools that allow us to measure MT dynamics, as well as the dynamics of other key cellular proteins, and measure how they change after drug treatment in living cells in real time (Matov, 2024f). Because tubulin inhibitors are already widely used for the treatment of many solid tumors, elucidating the mechanisms behind MT organization in cancer cells of epithelial origin could identify putative targets. To best benefit individual patients, one will ultimately need a procedure to test the drugs *ex vivo* within a time frame suitable for the context of the patient's treatment plan. In the Clevers lab (Hubrecht Institute, NL), the pioneers of organoid technology, there has been an organoid-based test for cystic fibrosis for over a decade that delivers results in about 2 weeks (Clevers and Bender, 2015). A major goal is to transfer this technology to oncology, which would allow for the dynamic analysis of MTs in patient-derived cancer cells during *ex vivo* treatment and, in doing so, enable the clinical use of MT dynamics as a biomarker for drug susceptibility.

## RESULTS

### Mitosis in Breast Cancer Cells

Aggressive forms of BC, such as receptor triple-negative cancers and estrogen receptor (ER)-positive cancers of the luminal B molecular subtype, are associated with worse patient outcomes (Maqbool et al., 2022). Although initially effective in reducing tumor burden for some BC patients, resistance to MT-targeting systemic therapy is universal and the cell biological profiling of patient-derived cells *ex vivo* is not utilized as a rationale for identifying drug-sensitive and drug-resistant patient populations prior to initiating therapy. The signaling molecules that interact with MTs, as well as the multiple effects on signaling pathways that destabilize or hyper-stabilize MTs, indicate that MTs are likely to be critical to the spatial organization of signal transduction. MTs are also affected by signaling pathways and this contributes to the transmission of signals to downstream targets (Gundersen and Cook, 1999). Many mechanistic models have been proposed to describe drug resistance to tubulin inhibitors (Orr et al., 2003), the most common of which are: (i) resistance due to membrane-pump overexpression (Robert, 1999), (ii) resistance due to tubulin-isotype composition (Haber et al., 1995; Kanakkanthara et al., 2011; Ranganathan et al., 1998), and (iii) resistance due to mutations in drug binding sites (Giannakakou et al., 2000; Giannakakou et al., 1997) as well as regulatory microRNA (Kanakkanthara and Miller, 2012). Evidence for these models stems largely from work with tissue culture cancer cell lines. Because these lines fail to adequately represent the genetic diversity found in the patient population, it is unclear how these findings relate to resistance to tubulin inhibitors in real tumors from patients.

To investigate the differences between receptor-negative and receptor-positive BC, we measured MT ends (plus tips) dynamics in both interphase (Matov, 2025a) and mitotic (Matov, 2025b) cells in six BC cell lines. Our measurements showed two main characteristics of the TNBC cells. In interphase cells, the EB1ΔC-2xEGFP speeds were significantly faster for the MDA-MB-231 cell line, with about 6 µm/min as a median value, than in receptor-positive cells (Matov, 2024c; Matov et al., 2015). The accelerated MT polymerization rates in MDA-MB-231 cells during interphase likely shorten interphase and accelerate cell cycle progression and mitotic entry, thus increasing their proliferation rates. In all mitotic cells, the EB1 speeds measured at the metaphase plate were around 21 µm/min (Fig. 1A shows a large mitotic spindle in a receptor-positive BC cell). We did not measure any significant difference in the MT flux (Fig. 2B) in either receptor-negative or receptor-positive cells during metaphase (n = 10, per cell line) for any of the six cell lines. This indicates that the mechanisms related to differential regulation might be related to the interactions with cytoplasmic molecular motors of the astral and non-kinetochore MTs. Such MT bundles are not tightly anchored, unlike the kinetochore MTs, and their behavior is dependent on the activity of cytoplasmic dynein.

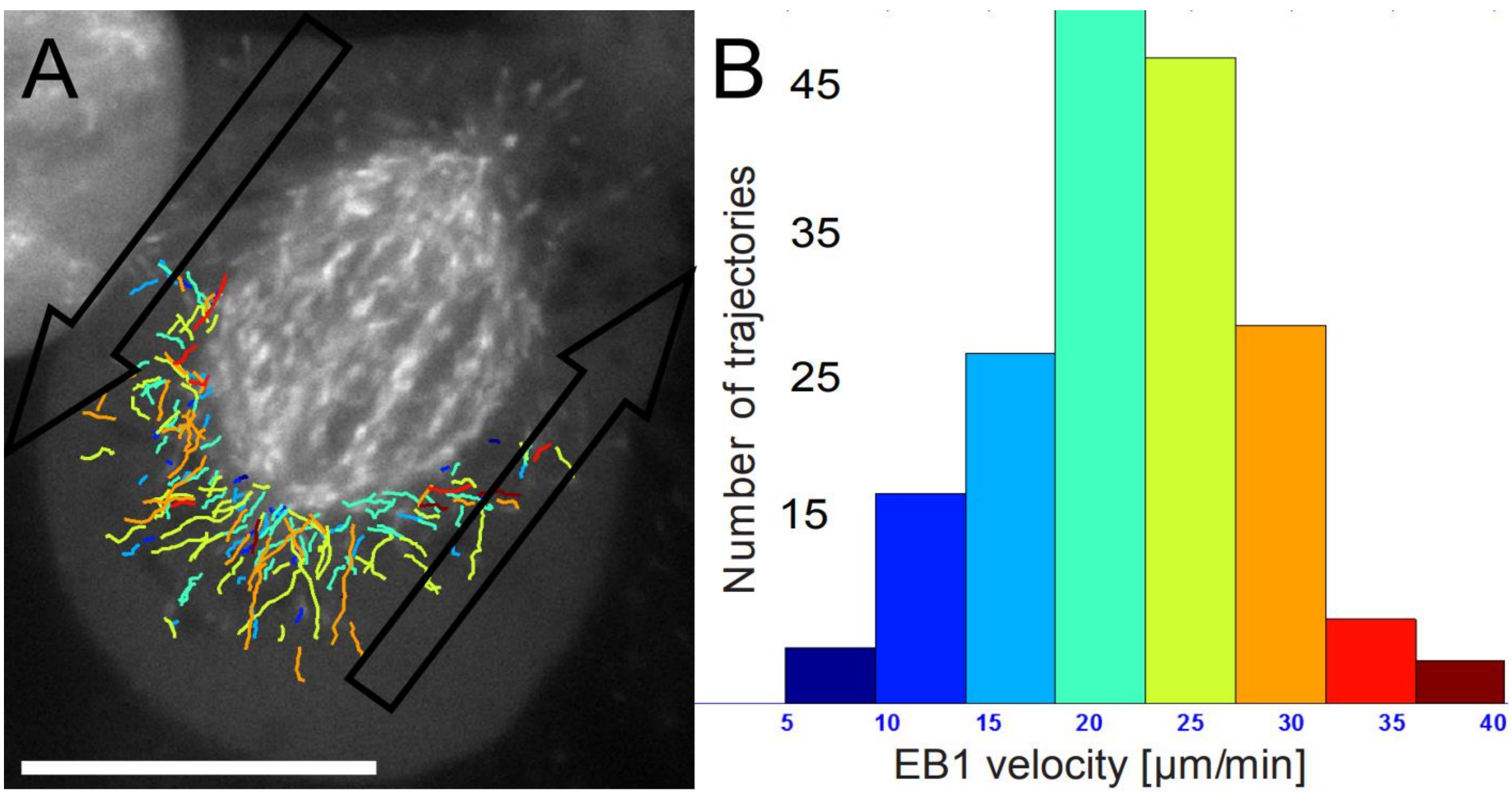

**Figure 1. Mitotic spindle and MT dynamics in a dividing receptor-positive (HCC-1428) BC cell.** (A) MT tips are labeled with EB1ΔC-2xEGFP, imaged for a minute (with an acquisition rate of two images per second), and computationally tracked (Yang et al., 2005). Scale bar equals 40 µm. The mitotic spindle exhibits antiparallel MT flux (Matov, 2024b) and the main two directions are marked with black arrows on the figure. The figure displays, as an overlay, the EB1 trajectories for a specific area of the spindle, namely, only the lower half (below the metaphase plate) and MTs emanating away from the spindle body, the so-called astral MTs. The color-coding represents EB1 speeds and colder colors correspond to lower speeds, and warmer colors correspond to faster speeds. The spindle appeared stationary and EB1 speeds were associated with MT polymerization rates. (B) Histogram of EB1 speed probability density function of the trajectories shown in Fig. 1A. The average speed is 20.5 ± 6.2 µm/min. The maximal EB1 speed is about 40 µm/min. The speed distribution is unimodal and normal.

In this context, what we found remarkable was the rocking pendulum-like motion in many of the mitotic spindles in MDA-MB-231 cells (Fig. 2A and Vid. 1). First, in interphase cells, the MDA-MB-231 cells exhibit the fastest MT polymerization dynamics at about 19 µm/min. Second, in metaphase, MDA-MB-231 cells have the largest spindles. While most spindles in the remaining five BC cell lines remain stationary until anaphase, many of the MDA-MB-231 spindles rotate at up to an angle of ±40º (Matov et al., 2015) to the left and right of the polar spindle axis (Vid. 1 and Vid. 2). This rapid rocking motion over an 80º angle is reflected in the measurements for the EB1 velocity (Fig. 2B). Altered MT dynamics endow cancer cells with both survival and migratory advantages (Mitchison, 2012), which can be correlated with drug resistance (Gonçalves et al., 2001). We observed similar rocking motion in dividing cells from the other TNBC cell lines as well.

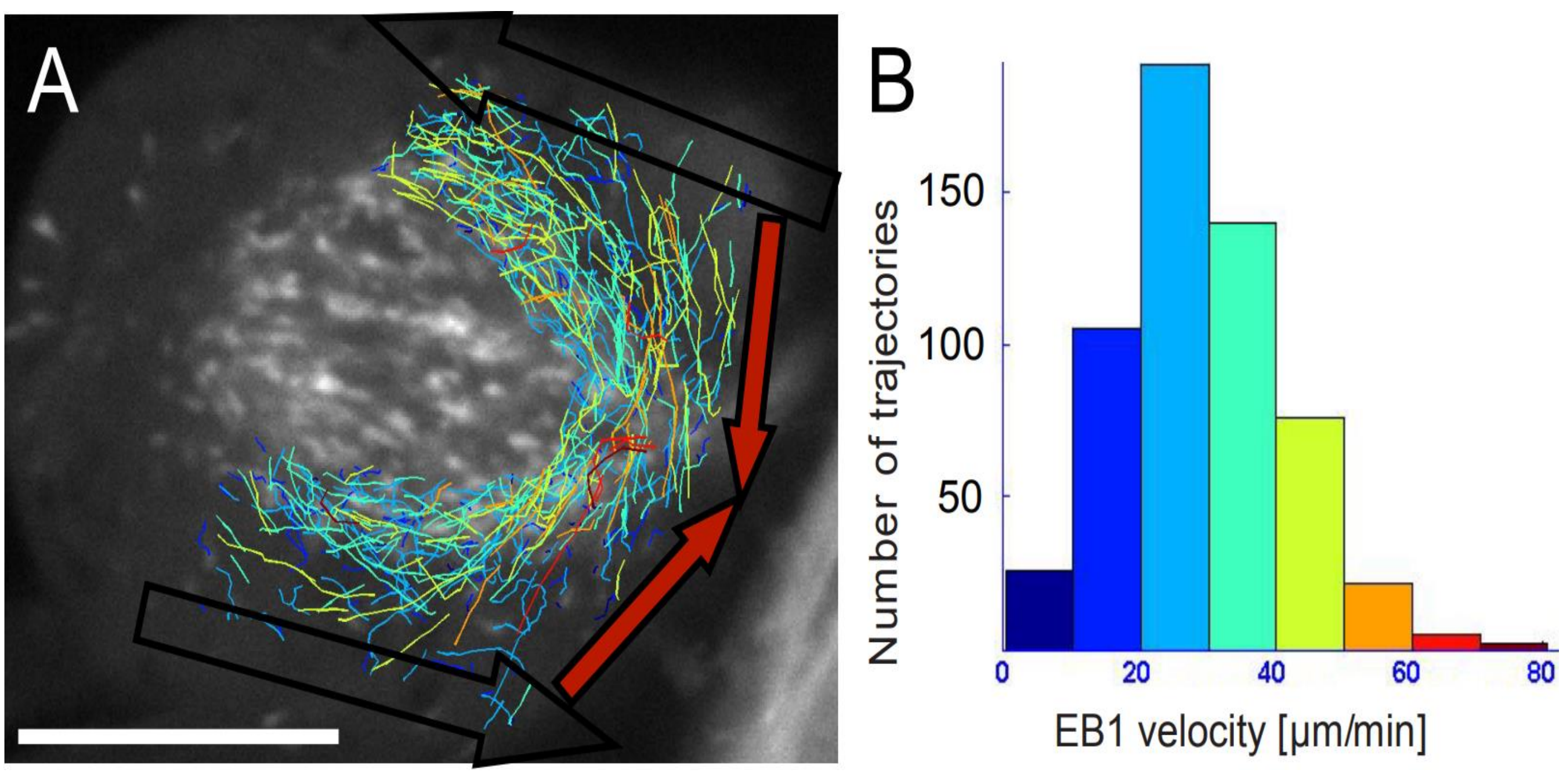

**Figure 2. Mitotic spindle and MT dynamics in a dividing receptor-negative (MDA-MB-231) BC cell.** (A) MT tips are labeled with EB1ΔC-2xEGFP, imaged for a minute (with an acquisition rate of two images per second), and computationally tracked (Yang et al., 2005). Scale bar equals 40 µm. The mitotic spindle exhibits antiparallel MT flux (Matov, 2024b) and the main two directions are marked with black arrows on the figure. In addition, the spindle exhibits rapid rotational motions and rocks to the left and to the right; the directions of motion of the spindle are marked with red arrows. The figure displays, as an overlay, the EB1 trajectories for a specific area of the spindle, namely, only the right half (to the right of the metaphase plate) and MTs emanating away from the spindle body, the so-called astral MTs. The color-coding represents EB1 speeds; colder colors correspond to lower speeds, and warmer colors correspond to faster speeds. The EB1 speeds were associated with the contribution of both MT polymerization rates and spindle rocking. Because of the extensive motion of the spindle, part of it is obscured by the trajectories overlaid on the image. To observe the rocking motions in the spindle, please see Vid.1. (B) Histogram of EB1 speed probability density function of the trajectories shown in Fig. 1A. The average speed is 29.7 ± 12.1 μm/min. The maximal EB1 speed is about 80 μm/min. The speed distribution is unimodal and skewed with a few very fast-moving EB1 comets.

## Microtubule Regulation

There are about 70 MT-associated proteins (MAPs) that regulate various stages of MT homeostasis (Lyle et al., 2009a; Lyle et al., 2009b). Variations in the expression levels of these genes have been linked to whether or not a tubulin inhibitor is able to kill cancer cells (Albrethsen et al., 2014; Berges et al., 2014; Devred et al., 2008; Le Grand et al., 2014; Li et al., 2014; Malesinski et al., 2015; O'Rourke et al., 2014; Thomas et al., 2015; Xie et al., 2015). There are also about 80 proteins at the kinetochore during cell division (Musacchio and Salmon, 2007). Of all solid tumors, BC is the disease with the highest number of approved tubulin inhibitors, consisting of seven stabilizing and destabilizing MT-targeting drugs, suggesting variability in patient response. It is unclear whether tubulin inhibitors kill mitotic or interphase cells. It is likely that mitosis is not the target (Field et al., 2014; Kitagawa, 2011; Komlodi-Pasztor et al., 2011; Komlodi-Pasztor et al., 2012; Komlodi-Pasztor et al., 2013; Mitchison, 2012; Ogden et al., 2013; Weaver, 2015; Wissing et al., 2013) and that the mechanisms by which tubulin inhibitors kill cancer cells in interphase are either absent in tissue culture or masked by high proliferation rates of cultured cells in a Petri dish. Recently developed compounds, which bind to MT plus ends and are designed to induce mitotic arrest (Kapoor et al., 2000), have failed in clinical trials. Drugs like paclitaxel and eribulin (Smith et al., 2010) induce lower mitotic arrest than ispinesib, which failed clinical trials (Dumontet and Jordan, 2010), further highlighting our gap in knowledge in this matter (Topham et al., 2015). These data suggest that, rather than mitotic arrest, a different mechanism, coupled to MT binding in interphase cells, might be essential for the antitumor effect *in vivo* (Zhang et al., 2015). In interphase, a very interesting mechanistic question relates to the ability of different drugs binding to the same site to affect dynamics differentially (Azarenko et al., 2014; Dumontet et al., 2009; Jordan and Wilson, 2004; Kamath et al., 2014; LaPointe et al., 2013; Ngan et al., 2000; Okouneva et al., 2003; Risinger et al., 2014).

There are four canonical parameters describing MT dynamics: MT growth speed, MT shortening (or shrinking) speed, catastrophe frequency (which occurs when a growing MT switches to a phase of shortening – it equals one over the time a MT spends growing) and rescue frequency (oppositely, this parameters relates to an event when a shortening MT switches to a phase of growth – it equals one over the time a MT spends shrinking) (Fig. 3, adapted from (Cassimeris et al., 1988)). As a fifth parameter, the time of pausing has commonly been accepted as the time a MT spends neither growing nor shortening. In a scenario in which we include pausing as a third MT state besides growth and shortening, there will be additional transitional frequency parameters back and forth between growth-to-pause as well as back and forth between shortening-to-pause. Table 1 provides examples of regulators of concrete aspects of MT behavior (Holmfeldt et al., 2009; Sousa et al., 2007). Functional redundancy between specific assembly factors in the chromatin pathway suggests that proteins/pathways commonly viewed as essential may not have unique functions, which is important in our quest to target the mitotic spindle in cancer.

**Interphase in Breast Cancer Cells**

The cell spends most of its life in interphase. During this time, the MTs are radially stretched from the nucleus toward the cell edge. Linear motion of proteins on MTs during interphase is common in all eukaryotic cells, such as MT plus-end-directed trafficking of proteins or any cargo transported on

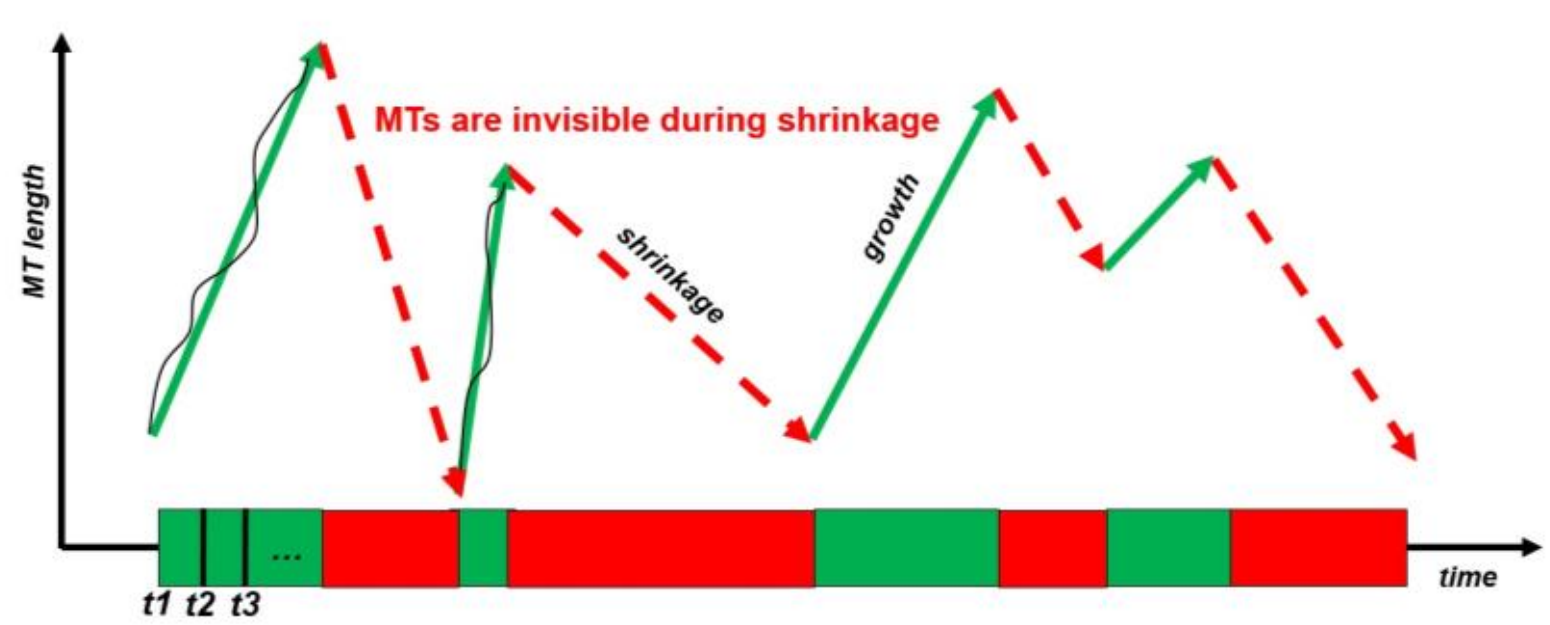

**Figure 3. Evolution of a MT length in time.** Using labeled EB1 proteins, we can directly measure only two of the MT dynamics parameters, MT growth speed and times. Since there is no molecular marker for MT pausing nor MT depolymerization (i.e., one cannot visualize the MTs during these stages), we infer computationally the remaining parameters by assigning multiple visible MT growth tracks to the same trajectory, as depicted in the figure (adapted from (Cassimeris et al., 1988)). Every switch from one phase to another, and changes in the behavior during these phases, is regulated by the genes listed in Table 1.

MTs. Computational analysis of this motion will allow secondary screening of novel tubulin inhibitors or targeted therapies affecting MAPs. In this context, we performed computational analysis of four BC cell lines before and after the addition of six tubulin inhibitors, three stabilizers and three destabilizers (Table 2). Because tubulin inhibitors are used in the treatment of multiple solid tumors and it is our view that more of them will be approved to treat a variety of tumors of epithelial origin, such as prostate cancer (PCa) and colorectal cancer, we envisaged testing FDA-approved and clinically used MT-stabilizing drugs, such as the taxanes, paclitaxel and docetaxel, as well as some of the MT-destabilizing drugs, such as the vinca alkaloids, vinblastine and vincristine. In addition, we included in our screen ixabepilone, another MT stabilizer used to treat breast tumors resistant to taxanes (Puhalla and Brufsky, 2008), and eribulin, a MT destabilizer that acts at growing MT ends by a distinct mechanism (Smith et al., 2010). To test different concentrations of these drugs, we performed a pilot study on a panel of five breast cell lines (four BC cell lines and one cell line from normal breast epithelial cells) and treated them with the six tubulin inhibitors. The four BC cell lines we used represented common BC molecular types – Hs578-T (receptor-negative), MDA-MB-231 (receptor-negative), MCF-7 (ER-positive), SK-BR-3 (human epidermal growth factor receptor 2 (HER2)-positive) as well as the normal breast epithelium cells, MCF-10A (Matov, 2024a). We performed titration experiments ranging from 0.1 to 100 nM concentrations and established that very low drug concentrations of 0.5 nM for MT-stabilizers and 0.1 nM for MT-destabilizers (which have higher potency) allow the cells to remain viable for several days (over a week) after treatment. Because the focus of our analysis is to identify a subset of MT parameters differentially affected by the MT-targeting drugs, higher concentrations, which collapse the cytoskeleton within hours of treatment, will be inappropriate and of questionable translational value considering the pharmacokinetics of many of these drugs in the body (Zasadil et al., 2014).

| Parameter name | MT-regulating proteins | Drug resistance /sensitivity | MT polymerization/ depolymerization activity |
|---|---|---|---|
| Growth speed | CKAP5<br>MAP4<br>TPX2<br>MAP2/Tau<br>CLIP170 | CKAP5<br>MAP4<br>TPX2<br>MAP2/Tau<br>CLIP170 | The rate at which MTs add new dimers, that is, polymerize |
| Shrink speed | KIF2C<br>KIF18B<br>TPX2 | KIF2C<br>TPX2 | The rate at which MTs remove dimers, that is, depolymerize |
| Pause times | CLASP<br>VHL | VHL | The time MTs spend stalled, that is, neither growing nor shortening |
| Catastrophe frequency | KIF2C<br>Op18<br>Katanin | KIF2C<br>Op18 | Frequency of switch from MT growth to shortening |
| Rescue frequency | CLASP<br>MAP2/Tau<br>Katanin | MAP2/Tau | Frequency of switch from MT shortening to growth |

**Table 1. Partial list of MT dynamics parameters (Matov et al., 2010).** Every aspect of MT dynamicity in living cells is controlled (Bowne-Anderson et al., 2015) by multiple MT-related genes (Lyle et al., 2009a; Lyle et al., 2009b), such as Op18 and MAP2, whose up- or down-regulation is linked at the same time to resistance to tubulin inhibitors (Inoué and Salmon, 1995; Lyle et al., 2009a; Lyle et al., 2009b; Xie et al., 2015). The full lists of tubulin inhibitors and MT parameters are longer and the regulation is more complex.

Our analysis showed that the receptor-negative Hs578-T cells were most susceptible to treatment with paclitaxel (see Table 2, highlighted in red - changes in the values of 10 MT parameters after drug treatment were statistically significant) and ixabepilone (Table 2, highlighted in red - changes in the values of 10 MT parameters after drug treatment were statistically significant). Conversely, according to our measurements, the receptor-negative MDA-MB-231 cells were resistant to all six drugs. The ER-positive MCF-7 cells were susceptible to treatment with paclitaxel (Table 2, highlighted in red - changes in the values of nine MT parameters after drug treatment were statistically significant) only. The HER2-positive SK-BR-3 cells were most susceptible to treatment with ixabepilone (Table 2, highlighted in red - changes in the values of eight MT parameters after drug treatment were statistically significant) and a vinca alkaloid, vinorelbine (Table 2, highlighted in red - changes in the values of eight MT parameters after drug treatment were statistically significant). Overall, the study showed that there is variability in drug resistance between TNBC cells and provided a rationale for the selection of chemotherapy in the HER2 overexpression setting. According to our computational analysis (Matov et al., 2010), ixabepilone (Goodin, 2008a), a semi-synthetic analog of epothilone B (Goodin, 2008b) would be the most efficient drug to treat the SK-BR-3 (Fogh et al., 1977) cells. The implications resulting from this determination are that ixabepilone (and vinorelbine) could be used as a first-line chemotherapy for HER2-positive breast tumors and also, in some cases, for TNBC tumors. This selection would spare patients genetically predisposed to not respond to the toxicity of taxanes. We reached a similar conclusion regarding the treatment of the most resistant cells, MDA-MB-231. For these cells, ixabepilone (see Table 2; changes in the values of four MT parameters after drug treatment were statistically significant) and vinorelbine (Table 2; changes in the values of four MT parameters after drug treatment were statistically significant) affected more parameters than paclitaxel, which approved as a first-line chemotherapy for metastatic BC. In the case of MDA-MB-231 cells, only up to four MT parameters were significantly affected, which suggests a limited impact of treatment with tubulin inhibitors.

Taken together, the results indicate that an assay based on the analysis of living patient-derived cells may offer improvements to the approach established in the current standard of care. The results also suggest the utilization of a data-driven algorithm for the identification of efficacious regimens. Further, they also provide an approach for the selection of off-label tubulin inhibitors, which in some cases induce long-lasting remission even in very advanced disease stages (Linn et al., 1994). We envisage a system in place that utilizes pairs of patient normal tissue cells and neoplastic tissue cells for which cytoskeletal reorganization in disease is assessed *ex vivo*. This way, like in a game of Tetris, drugs may be selected for their ability to reverse the full spectrum of disease-induced changes in MT dynamics in anticipation of maximal treatment efficacy. Our approach might be impactful beyond BC, also in the treatment of other solid tumors. Currently, tubulin inhibitors are not approved for the treatment of many tumors of epithelial origin based on clinical trials, which require the vast majority of patients to respond to treatment, whereas in some cancers the response is limited to a fraction of the patients – however, some patients, even if a few, do respond.

**Mitosis in Patient Cells**

Even though the overall regulation is a complex process, three nucleation/stabilization factors (Groen et al., 2009), TPX2, γ-tubulin, and CKAP5/XMAP215 are required for MT polymerization and bipolar spindle assembly around chromatin beads in an *in vitro* system. Depletion of TPX2 was partially rescued by inhibiting KIF2C/MCAK or by adding EB1 or CKAP5/XMAP215. Depletion of CKAP5/XMAP215 or γ-tubulin was partially rescued by adding CKAP5/XMAP215, but not by adding other factors. This demonstrates that the functioning of the spindle requires a minimal level of multiple MT regulators. In this context, aberrant MT dynamics in cancer can be exploited by targeting the specific genes responsible for generating them. It is conceivable that (i) modulation of multiple MT regulators will reverse resistance and sensitize cancer cells to tubulin inhibitors (Matov, 2024d) and (ii) further modulation will eventually make the cells apoptotic (Matov, 2024b). Differential expression of MT-regulating genes affects specific parameters of MT dynamics, thus determining sensitivity to treatment. Combining personalized tubulin inhibitor chemotherapy with a customized list of cancer-specific targets within the MT-regulating genes based on the techniques described might offer therapeutic plans resulting in both low toxicity and low disease relapse rates.

In previous work (Harkcom et al., 2014), we measured an increase in MT dynamicity in a BC MCF-7 cell line after treatment with nicotinamide adenine dinucleotide ($NAD^+$), an endogenous small molecule that has effects on diverse processes, including obesity, lifespan, and cancer, by tracking the motion EB1 comets. We report new measurements indicating that cells pretreated with $NAD^+$ are resistant to vinblastine. Parental MCF-7 cells no longer had polymerizing MTs after 1-hour treatment with 5 nM vinblastine. To measure the effects on MT dynamics, we reduced the expose to half an hour and measured an attenuation of MT polymerization rates of 3 µm/min (Table 3). However, if the cells were treated with 5 mM $NAD^+$ and then exposed to vinblastine, the reduction was 2 µm/min. Further, 10 mM $NAD^+$ additionally limited the ability of vinblastine to impact MT turnover rates to a reduction of just 1 µm/min. These results suggest that lowering the levels of $NAD^+$ in ER-positive tumors might sensitize them to treatment with vinca alkaloids.

| Cell line and drug treatment | Growth Speeds *μm/min* | Pause Speeds *μm/min* | Growth Times *second* | Pause Times *second* | Growth Segm. Speeds *μm/min* | # Pause Per Segm. *A.U.* | % T. in Pau. *A.U.* | Growth Segment Times *second* | Shrink Speeds *μm/min* | Shrink Times *second* | Pro. MT Switches to Shrink *A.U.* | % Time Spent in Shrink *A.U.* |
|---|---|---|---|---|---|---|---|---|---|---|---|---|
| Hs578-T | 11.21 | 4.53 | 7.32 | 11.44 | 7.02 | 1.38 | 43 | 32.55 | 8.07 | 16.08 | 19.34 | 13.87 |
| Hs578-T PTX | 12.39* | 5.28* | 6.99* | 10.53* | 7.77 | 1.30* | 41 | 28.99* | 9.84* | 16.69* | 26.53* | 23.30* |
| Hs578-T DTX | 10.68 | 4.41 | 7.80 | 10.40 | 6.79 | 1.36 | 41 | 32.32 | 8.61 | 16.15 | 30.81* | 23.73* |
| Hs578-T iX | 15.35* | 6.86* | 7.78* | 9.96* | 10.18 | 1.35 | 40* | 30.43* | 12.82* | 16.64* | 29.23* | 25.77* |
| Hs578-T ER | 10.72 | 4.48 | 7.29 | 11.01 | 6.94 | 1.42 | 44 | 31.65 | 8.84 | 15.66 | 27.98* | 21.99* |
| Hs578-T VNL | 12.92* | 5.22 | 6.72* | 11.48 | 8.11 | 1.41 | 46* | 31.24 | 9.79* | 15.70 | 26.01* | 21.28* |
| Hs578-T VNC | 12.08 | 4.74 | 7.16 | 11.38 | 7.17 | 1.37 | 44 | 31.18 | 9.78 | 15.59 | 30.46* | 24.44* |
| MDA231 | 15.83 | 7.84 | 6.10 | 10.30 | 10.24 | 1.27 | 46 | 25.14 | 14.47 | 16.94 | 44.16 | 48.02 |
| MDA231 PTX | 16.29 | 7.42 | 6.39 | 10.41 | 10.42 | 1.32* | 47* | 27.12 | 13.66 | 17.23 | 43.34 | 49.15 |
| MDA231 DTX | 11.79 | 5.21 | 6.32 | 10.39 | 7.53 | 1.28 | 45 | 26.37 | 9.71 | 17.30 | 40.19 | 44.67 |
| MDA231 iX | 19.35* | 8.99* | 6.29 | 10.06 | 12.19 | 1.26 | 47* | 25.92 | 16.43* | 16.87 | 44.87 | 50.32 |
| MDA231 ER | 15.17 | 6.33 | 5.72 | 11.16 | 8.55 | 1.22 | 49 | 25.63 | 12.88 | 17.26 | 43.73 | 48.32 |
| MDA231 VNL | 19.74* | 8.75 | 5.86 | 9.91 | 12.42 | 1.29 | 45 | 25.50 | 16.66* | 17.10 | 36.43* | 40.76* |
| MDA231 VNC | 14.83 | 6.79 | 6.15 | 10.85 | 9.16 | 1.32 | 47 | 28.23 | 13.72 | 16.12 | 40.60 | 42.73 |
| MCF-7 | 12.54 | 5.52 | 7.55 | 11.14 | 8.09 | 1.32 | 41 | 31.36 | 9.69 | 15.85 | 29.63 | 22.72 |
| MCF-7 PTX | 10.19* | 4.42* | 7.42 | 10.46* | 6.95* | 1.28* | 42 | 28.78* | 7.25* | 16.53* | 34.58 | 30.35* |
| MCF-7 DTX | 9.59 | 4.24 | 7.99 | 11.09 | 6.43 | 1.33 | 41 | 31.57 | 8.11 | 15.90 | 34.66 | 27.77 |
| MCF-7 iX | 12.52 | 5.57 | 8.05* | 10.39 | 8.54 | 1.30 | 41 | 30.16 | 9.68 | 16.06 | 33.96 | 27.17 |
| MCF-7 ER | 10.33 | 4.55 | 7.37 | 10.81 | 6.88 | 1.32 | 43 | 30.44 | 7.91 | 16.90 | 31.75 | 27.33 |
| MCF-7 VNL | 12.57 | 5.36 | 7.03* | 10.68 | 8.22 | 1.33 | 43 | 29.67 | 9.02 | 16.34 | 30.52 | 25.71 |
| MCF-7 VNC | 9.22* | 4.01 | 7.05 | 10.85 | 5.96* | 1.32 | 42 | 29.63 | 6.87* | 15.95 | 29.39 | 26.48 |
| SK-BR-3 | 11.30 | 4.95 | 6.16 | 12.20 | 6.69 | 1.23 | 49 | 27.89 | 8.59 | 15.62 | 36.30 | 29.50 |
| SK-BR-3 PTX | 11.88 | 5.17* | 6.05 | 11.62 | 7.31* | 1.14* | 48 | 25.41* | 8.78 | 16.51 | 42.30* | 41.51* |
| SK-BR-3 DTX | 12.18 | 5.30 | 5.79 | 11.76 | 7.45 | 1.22 | 51 | 26.40 | 10.22 | 16.63 | 49.45* | 53.33* |
| SKBR3 iX | 16.14* | 7.28* | 5.91 | 11.55 | 9.96 | 1.18* | 49 | 25.85* | 13.27* | 16.35* | 50.66* | 51.95* |
| SKBR3 ER | 12.69 | 5.42 | 5.78 | 12.40 | 7.63 | 1.24 | 50 | 27.11 | 10.38 | 16.64 | 48.23* | 48.66* |
| SKBR3 VNL | 15.01* | 6.84* | 5.48* | 11.89 | 9.53 | 1.15* | 50* | 24.97 | 11.85* | 16.14 | 46.86* | 49.49* |
| SKBR3 VNC | 11.68 | 5.24 | 5.41 | 11.91 | 7.21 | 1.19 | 51 | 25.80 | 10.72 | 16.02 | 47.50* | 51.85* |

**Table 2. Quantitative measurement of MT dynamics in four BC cell lines treated with six tubulin inhibitors.** Computational analysis of MT dynamics was performed using ClusterTrack (Matov et al., 2010). The cell lines consisted of triple-negative cells (Hs578-T), triple-negative cells (MDA-MB-231), ER-positive cells (MCF-7), and HER2-positive cells (SK-BR-3), representing the most common types of breast cancers exhibited by metastatic patients. The MT inhibitors used were ixabepilone (iX), docetaxel (DTX), and paclitaxel (PTX) (MT stabilizers; treatment at a concentration of 0.5 nM), eribulin (ER), vinorelbine (VNL), and vincristine (VNC) (MT destabilizers; treatment at a concentration of 0.1 nM). For each of the 28 conditions, 20 movies were analyzed. Each movie consists of 120 frames taken 0.5 seconds apart. Asterisks (*) denote statistically significant shifts ($p < 0.05$) in MT parameter values, which were calculated on a per-cell basis as well as at the level of individual MTs. Observe the nonlinear effects of the drugs, as it is established that a high dose of paclitaxel stabilizes MTs, whereas, as we measured here, a low dose systematically increased MT polymerization rates in all cells except MDA-MB-231. The study also provided a rationale for the selection of chemotherapy in the HER2-positive setting, where, according to our computational analysis, ixabepilone would be the most efficient drug to treat the SK-BR-3 cells.

The 12 MT dynamics parameters (Matov et al., 2010) are as follows: (1) MT polymerization rates (growth speeds), (2) MT dynamics during trajectory gaps, when the EB1ΔC-2xEGFP molecular marker is not visible (pause/gap speeds), (3) MT polymerization periods (growth times), (4) periods of MT pausing and trajectory gaps, when the EB1ΔC-2xEGFP molecular marker is not visible (pause/gap times), (5) MT polymerization rates over multiple trajectories interrupted by periods when the EB1-EGFP molecular marker is not visible (growth segment speeds), (6) number of periods when the EB1ΔC-2xEGFP molecular marker is not visible within MT growth segments (number of pauses/gaps per growth segment), (7) percent of time MTs spend pausing within the growth segments (percent time in pause/gap), (8) periods of MT polymerization rates over multiple trajectories interrupted by periods when the EB1ΔC-2xEGFP molecular marker is not visible (growth segment times), (9) MT depolymerization rates (shrink/shortening speeds), (10) MT depolymerization periods (shrink/shortening times), (11) probability of MTs switching from polymerization to depolymerization, that is, undergoing catastrophe, at the end of a MT growth segment, and (12) percent of time MTs spend depolymerizing within composite MT trajectories (percent time in shrink/shortening).

The rates of polymerization, depolymerization, and pausing motion are measured in micrometers per minute. The duration of the periods of polymerization, depolymerization, and pausing are measured in seconds. The remaining parameters are measured in arbitrary units.

Polo-like kinase 1 (PLK1) is the principal member of the serine/threonine kinase family (Chiappa et al., 2022) and is overexpressed in TNBC (Ueda et al., 2019). PLK1 inhibition affects MT dynamics by suppressing polymerization rates (Brennan et al., 2008), which leads to spindle shortening. This effect on the regulation of mitosis can be reversed by the addition of paclitaxel (Brennan et al., 2008), which stabilizes MTs – in this scenario, at their minus ends. When MT plus-end-directed motors are suppressed (Matov, 2024b), MT sliding toward the spindle pole maintains MT flux, likely driven by cytoplasmic dynein (Gudimchuk and McIntosh, 2021), which suggests the use of combination therapies. One avenue to overcome resistance to tubulin inhibitors is to use a combination therapy with PLK1 inhibitors (Su et al., 2022).

| Treatment of MCF-7 cells | Reduction in EB1 speed |
|---|---|
| 5 nM vinblastine (VBL) only | 3 μm/min |
| 5 mM $NAD^+$, 5 nM VBL | 2 μm/min |
| 10 mM $NAD^+$, 5 nM VBL | 1 μm/min |

**Table 3. $NAD^+$ protects MCF-7 cells from the effects of vinblastine.** Cells pretreated with $NAD^+$ for 2 hours have a reduced sensitivity to treatment with 5 nM VBL.

The lack of methods for visualizing MT plus ends while they pause or depolymerize has been an obstacle to the full validation of the computational analysis of MT pausing and depolymerization (Matov et al., 2010). It would be informative to be able to quantify the dynamics of human MTs while they depolymerize, either at their plus or minus ends, by modifying proteins like Dam1 (Gardner and Odde, 2008), Ska1 (Ye and Maresca, 2013), NOD (Cane et al., 2013) or DDA3 (Jiang et al., 2012). Such direct measurements of a visible molecular marker would allow one to determine the secondary molecular mechanisms activated in drug-treated cancer organoids (Matov, 2025a) leading to the development of resistance and, thus, informing therapy and selecting combination therapies to which the disease would respond (Matov, 2024e).

To translate such results into clinically relevant models, one can transfer this technology to BC organoids (Sachs et al., 2018). Classification of trajectories can be accomplished by mimicking patient treatment in the organoids and accessing results of measurements from patient tumors which show whether the (i) tumors recede, that is, respond to treatment or (ii) grow or show no change, that is, exhibit drug resistance. Additionally, one can expect to be able to further subclassify the resistant signatures based on the mechanism behind them. These steps will be repeated and novel significant MT signatures will be recorded as treatment progresses, which will allow for the accumulation of large signature populations, thereby enabling statistical analyses. Furthermore, because organoid cultures are susceptible to molecular manipulations (Koo et al., 2013), one can reduce MT regulators expression by shRNA or CRISPR and determine whether this counteracts drug resistance and restores normal MT dynamics using mRNA-based drugs in a physiologically relevant system.

We conducted a preliminary investigation in colon cancer organoids (Fig. 4A), for which the organoid culture is better established (van de Wetering et al., 2015). Our measurements (Fig. 4B) indicate that these patient organoid cells have MTs polymerizing at rates characteristic for chromosomal instability (Ertych et al., 2014) and the EB1 comet speeds are significantly faster than those of the commonly utilized colon cancer cell lines when cultured in a monolayer, highlighting the importance of physiological 3D organoid model systems for the fidelity of the measurements. The cells in the organoid form a lumen and to the left a mitotic cell is visible (Fig. 4A). The mitotic spindle rotates rapidly (Vid. 3) similarly to the spindles in MDA-MB-231 cells and this suggests that this patient could be resistant to therapy. We observe and measure strong lateral rotation in the spindle (red arrows on Fig. 4A), which results in fast EB1 speeds (Fig. 4B and Vid. 4).

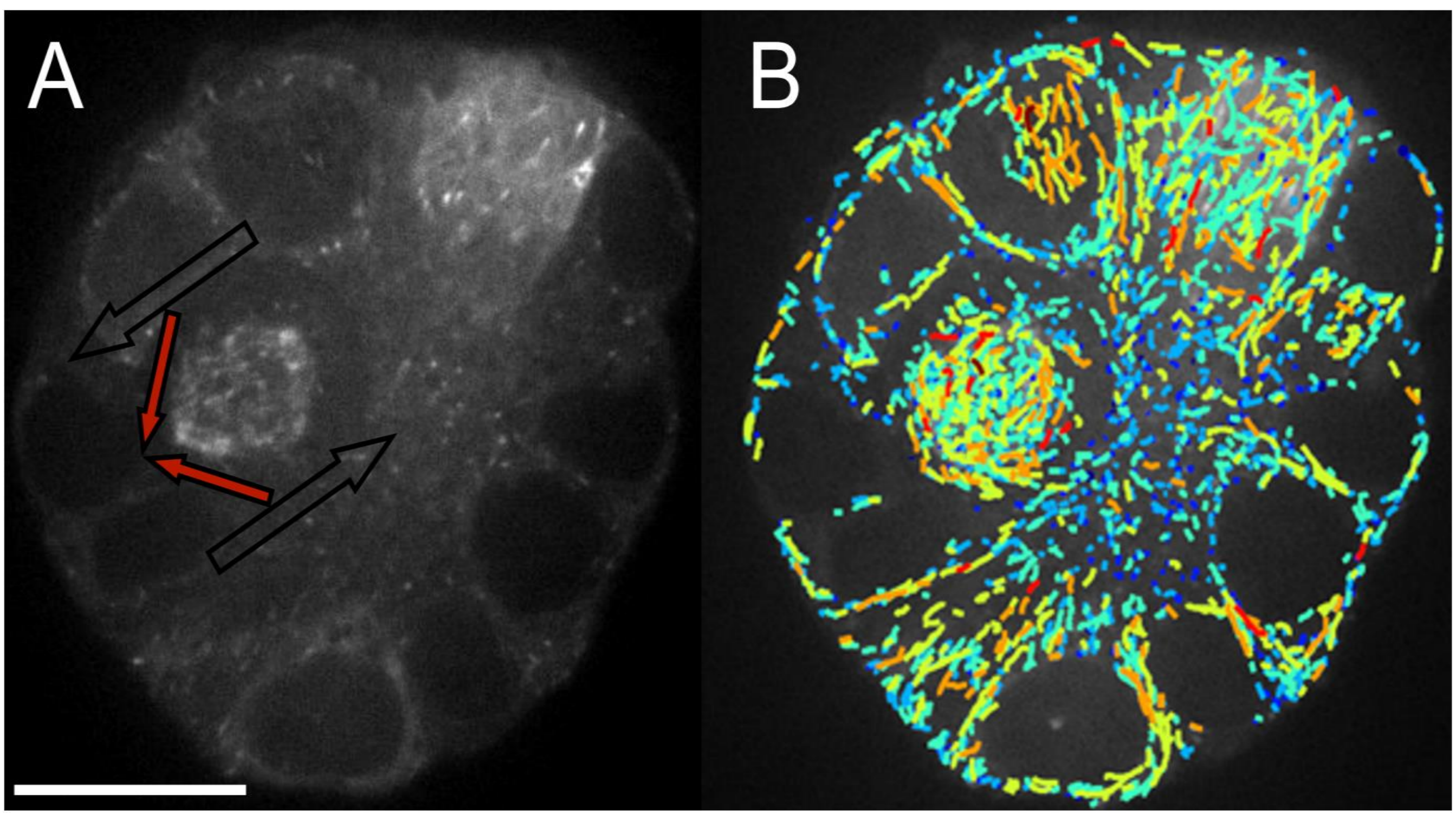

**Figure 4. Measurements of MT signatures in patient-derived cancer organoids.** (A) EB1ΔC-2xEGFP in stably expressed in colon cancer organoids (scale bar equals 70 µm). (B) Computer-generated MT growth tracks (Yang et al., 2005). The color-coding represents EB1 speeds; colder colors correspond to lower speeds, and warmer colors correspond to faster speeds. Note the mitotic spindle of a dividing cancer cell on the left side of the image. The average growth rate in this specimen is 19 µm/min, which is significantly faster compared with colon cancer cells cultured in a 2D monolayer, highlighting the importance of physiological 3D organoid model systems for the fidelity of the measurements.

## Interphase in Patient Cells

We expressed fluorescently tagged MT markers using lentivirus-mediated transduction and our MT dynamics analysis indicates differences between different PCa organoids (Fig. 5). Analysis of available RNA-Seq data shows intriguing differences in expression levels of MT regulators, and EB1 comet tracking indicates a correlation of decreased MT polymerization rates with increased Op18 (Fig. 5A) and KIF2C expression, providing an indication of a physiological mechanism of resistance in patient tumors. Organoid PCa1 shows high levels of MT-destabilizing factors, including Op18 (Matov et al., 2016) and had not undergone docetaxel chemotherapy (Gao et al., 2014), to which the tumor would likely have been resistant. We predict that we will find similar high Op18 expressing cases in taxane-resistant organoids established from BC patient cells. We could define a subtype of Op18-expressing tumors, and directly test how such patient cells and MT dynamics respond to different drugs compared with other organoids with low Op18 expression levels. Patient organoids overexpressing the tubulin-sequestering protein Op18 may be resistant to elevated taxane concentrations, but may still be sensitive to other tubulin inhibitors. This outcome would be clinically relevant and suggest a possible novel therapeutic strategy.

In particular, the approach can be applied for the *ex vivo* evaluation of the efficacy of mRNA-based drugs. The inset on Fig. 5A shows that the EB1 speeds in organoid PCa1 form a multi-modal distribution (the four peaks are marked with red lines). The reason behind this type of distribution may be that the activity of different genes regulating MT polymerization (Table 1), for example the expression levels of CKAP5/XMAP215, MAP2, and CLIP170, lead to the different modes in the speed distribution (Matov, 2025a). By knockdown or reconstitution of each of these genes and consequent analysis of the derivative cultures, we could delineate their precise contribution to the abnormal behavior of the MT cytoskeleton in disease. This approach can, thus, be utilized as a strategy for the selection of novel targets in the context of overcoming resistance to therapy in BC. Besides sensitizing drugs, our strategy could be utilized for the identification of putative targets within the MT transcriptome (Matov, 2024c).

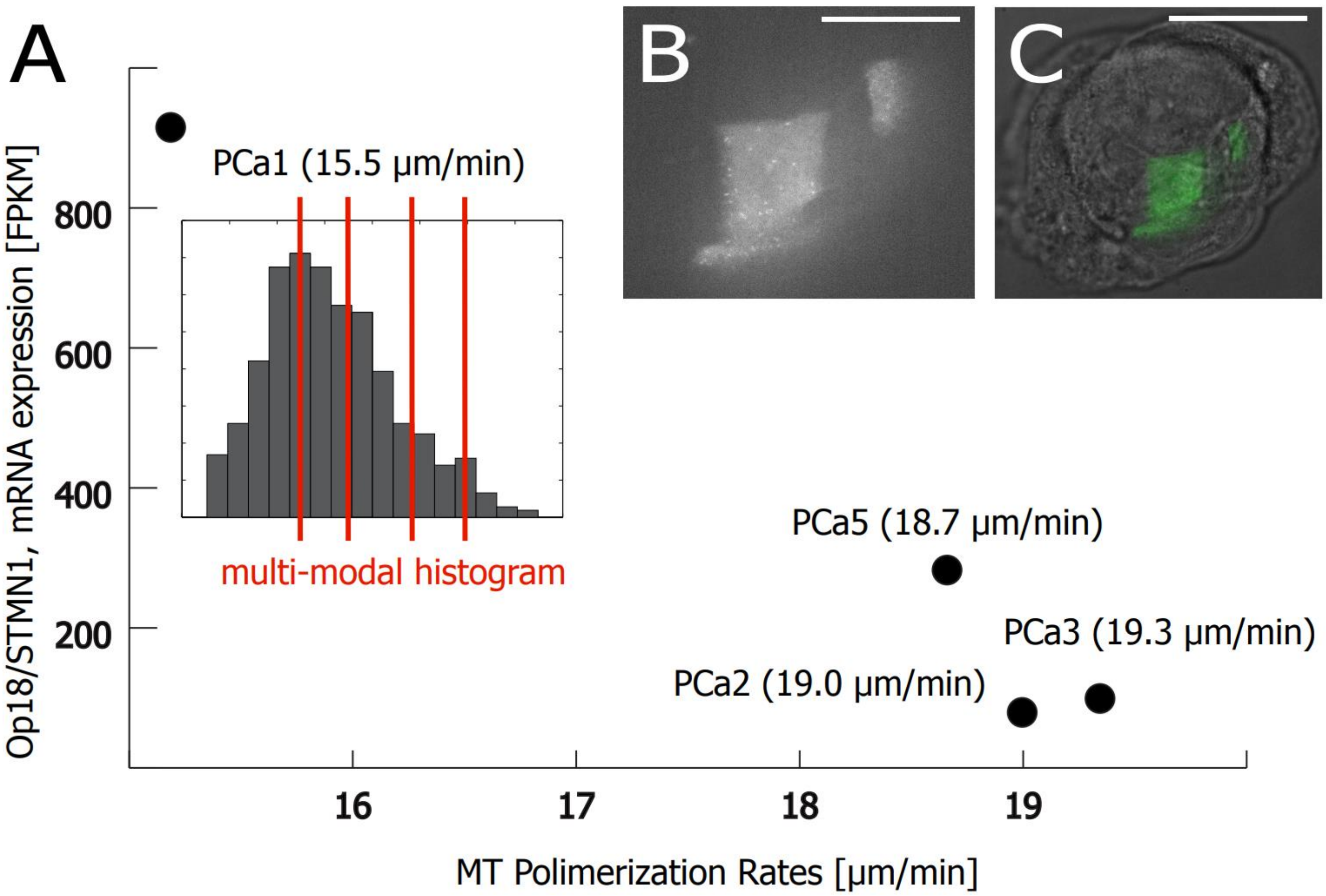

**Figure 5. RNA-Seq. data from the MSKCC metastatic castrate-resistant PCa organoids showing high expression of Op18 in PCa1 correlates with attenuated MT polymerization rates measured by motion analysis of EB1ΔC-2xEGFP comets.** (A) Numbers in brackets show preliminary MT growth rates from a limited number of cells measured (Yang et al., 2005); the inset shows the histogram of MT growth rates in PCa1 with a mean value of 15.5 µm/min; a multi-modal distribution may result from contributions of four or five genes that regulate MT growth. This analysis can be used to generate a shortlist of candidate genes involved in the regulation of MT dynamics for targeted therapy, such as RNA therapy, to sensitize resistant tumor cells. (B) Spinning disc confocal microscopy image of a patient organoid seeded and transduced with EB1ΔC-2xEGFP a day after radical prostatectomy; observe that only three cells in the round organoid are labeled. Scale bar equals 50 µm. (C) The same organoid as in (B), with the three EB1ΔC-2xEGFP expressing cells pseudo-colored in green and overlaid with the organoid imaged in transmitted light. Scale bar equals 70 µm.

## DISCUSSION

Besides spindle misorientation, loss of pVHL may cause chromosome instability and loss of astral MTs, which can be rescued by paclitaxel treatment (Thoma et al., 2009). In colorectal cancer cells, paclitaxel treatment could lead to overcoming chromosomal instability and successfully progressing through mitosis (Ertych et al., 2014). Chromosomal instability and impaired spindle checkpoint function are the result of reduced spindle assembly checkpoint mitotic arrest deficient 2 (Mad2) protein in *VHL*-defective cells (Thoma et al., 2009). Pathogenic levels of Rit1, a Ras-related GTPase, can suppress the spindle assembly checkpoint function by sequestering Mad2 from the mitotic checkpoint complex and promote chromosome segregation errors and aneuploidy (Cuevas-Navarro et al., 2021). The variability in regulation and redundancy in the activated pathways in disease demonstrate the need for quantitative live-cell microscopy methods that can determine the changes in morphology and dynamic organization within cells to predict clinical drug efficacy (Matov, 2024d).

The treatment of advanced BC is hampered by the emergence of drug resistance. We describe an approach for drug selection in BC, which is based on the computational analysis of live-cell microscopy of patient-derived cells *ex vivo*. It reveals differences in the response to tubulin inhibitors even in TNBC, for which the resistance to taxanes is not universal. Differences in the ability of drugs to kill cancer cells stem from the level of drug-target engagement, which cannot necessarily be predicted by genetic sequencing alone. Paclitaxel reduces the tension at the kinetochores, but that is insufficient to induce Mad2 accumulation on kinetochores; instead, kinetochores unattached to MTs consistently bind Mad2 (Waters et al., 1998) and progress to anaphase. Drug treatments that lead transiently to mitotic arrest rather than cell death, which is the basis for drug resistance, require combination regimens (Verissimo et al., 2016). All tumors of epithelial origin are susceptible to tubulin inhibitors, depending on the tumor molecular type (Mertens et al., 2023). In this context, quantitative live-cell microscopy (Matov, 2024e) may aid in anticipating whether a compound has the ability to affect the desired molecular target. It may also offer a strategy for targeting particular aspects of MT regulation to sensitize tumors to systemic therapy. The main difficulty in utilizing this technology is the need for off-line image analysis, which is tedious - a limitation that we have begun addressing with the development of real-time analysis systems (Matov, 2024e; Matov, 2024f), and the next challenge is clinical validation and approval.

## MATERIALS AND METHODS

### Image Analysis

All image analysis programs for the detection and tracking of comets, as well as graphical representation of the results were developed in Matlab and C/C++. The EB1 comet analysis method ClusterTrack used is described and validated in (Matov et al., 2010). The computer code is available for download at: https://www.github.com/amatov/ClusterTrackTubuline.

### Statistical Analysis

Statistical tests were performed in Matlab. We performed Kolmogorov-Smirnoff statistics (Kolmogorov, 1933) to compare higher order moments of the underlying MT parameter per track distributions and permutation *t*-test to extract information on shifts in median values of polymerization parameters over a population of cells. In brief, for each condition, multiple videos containing one or two cells were compared. Before merging the data for further statistical analysis, we determined the video-to-video variation in the distributions of MT parameters using the Kolmogorov-Smirnov test (Kolmogorov, 1933). Distributions were subsampled to 400 values to avoid hypersensitivity of the Kolmogorov-Smirnov statistics. Videos passing the consistency test were included in the final datasets. Statistical comparison of MT dynamics between different conditions was accomplished in two steps. First, we examined whether the different conditions would profoundly change the distribution of each parameter, which would indicate an entirely different state of MT regulation between conditions. To test this possibility, we aligned the distributions of compared conditions at their respective per-track mean value and then applied the nonparametric Kolmogorov-Smirnov test to determine whether the two distributions are identical. The distributions were randomly subsampled to 400 values. The subsampling was repeated 200 times, and the final p-values were derived by averaging. In a second step, we compared the shifts in parameter mean values between different conditions. To achieve this, we calculated the per-cell means for each parameter. Then, we tested the normality of the distribution of the means and applied a test to define the significance of parameter changes between conditions (mean ± SEM of cell-to-cell variation). This test assumes that the cell-to-cell variation is the single biggest source of variation in the mean value of a condition. None of the parameters of MT dynamics followed a Gaussian distribution. To determine the significance of differences in the mean values between parameters under different experimental conditions, we applied a non-parametric permutation *t*-test with 400 repetitions (Hesterberg et al., 2005), which compares the bootstrapped distributions of condition 1 versus condition 2. This test does not make any assumption about the characteristic of the tested distribution and, thus, is appropriate for application to distributions that are far from being normally distributed. In brief, for each condition 400 values are bootstrap-sampled from the data. In agreement with the central limit theorem (Laplace, 1810), the two bootstrapped distributions always follow a Gaussian distribution and, thus, could be analyzed for differences using a regular Student's *t*-test.

**Cell Culture**

All nine breast cancer cell lines were grown in DMEM (Cellgro, Mediatech) supplemented with 10% FBS (Atlanta Biologicals), incubated at 37°C in 5% $CO_2$. MT polymerization dynamics were visualized by lentivirus-mediated low-level expression of EB1ΔC-2xEGFP (Fig. 6A), a marker of growing MT plus ends that does not interfere with endogenous EB1 binding partners. The MDA-MB-231 TNBC cells were also grown as 3D cultures in Matrigel (Corning) with MT polymerization dynamics visualized by lentivirus-mediated low level expression of EB1ΔC-2xERFP.

Organoid medium was prepared according to modified protocols from the Clevers lab (van de Wetering et al., 2015) and the Chen lab (Gao et al., 2014). We expressed fluorescently tagged MT markers in the organoids by using lentivirus-mediated transduction (Fig. 6B) and had the best results infecting organoids in an exponential growth phase. To optimize imaging conditions, blasticidin selection was tuned so that only 20%-30% of organoid cells are GFP-labeled (Fig. 6C). This way, we achieved better contrast in imaging EB1 and MT dynamics.

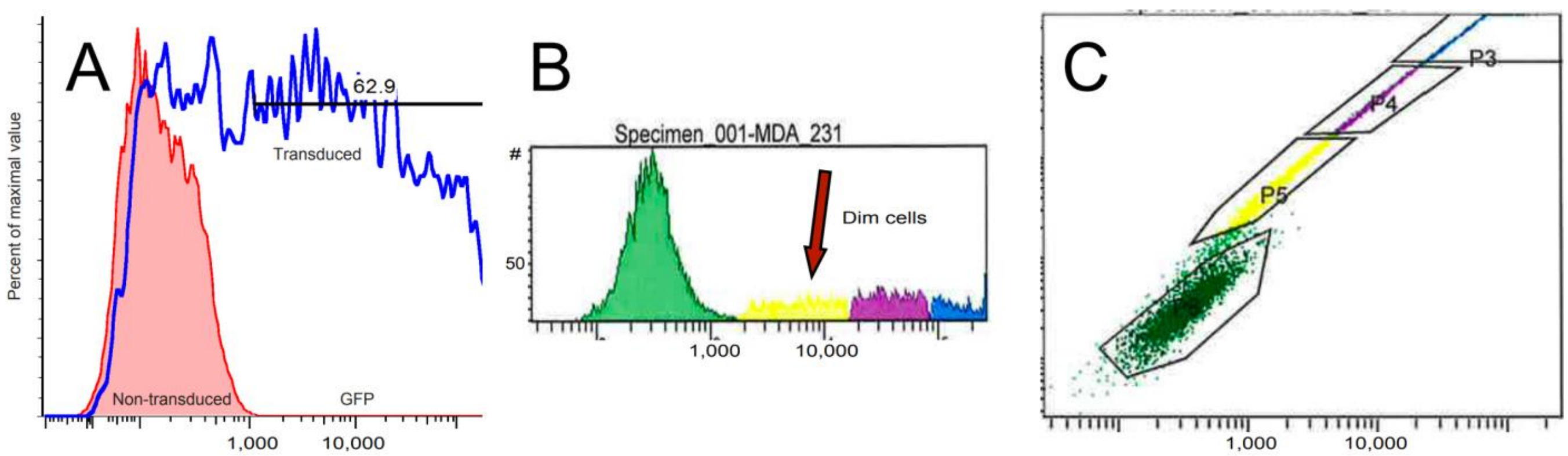

**Figure 6. Cell sorting of EB1ΔC-2xEGFP-expressing cells**. Following lentiviral gene transfer, 2,500,000 cells were from each cell line were sorted. After fluorescence-activated cell sorting (FACS), 40% of the cells were vital, according to DAPI labeling. (A) Among the infected cells, 62.9% were successfully transduced. (B) The transduced cells exhibited varying levels of GFP expression: 40% displayed low expression (dim, indicated in yellow – see the red arrow), 40% showed medium expression (in magenta), and 20% displayed high expression (bright, with labeling along the MT lattice that blocked MT dynamics, in blue) in GFP levels. We selected the dim cells (P5 in panel (C)), which provide the best image contrast. The sorted cells required about 10 days to recover and resume proliferating, and we collected EB1 images 12 days after infection. Although the MDA-MB-231 cells (data for which are shown in the figure) recovered well, some of the cell lines with more sensitive cells exhibited low vitality, leading us to conclude that this approach was not suitable for work with primary cells. We were able to optimize infection protocols and achieve significantly higher transduction rates, which then made FACS unnecessary.

To transduce organoids, we modified protocols from the Clevers lab (Koo et al., 2013) to adapt to the specifics of prostate organoid culture (such as the significant differences in proliferation rates in comparison to colon and rectal organoids). We found empirically that cells in mid-size organoids (60-100 µm in diameter) infect at much better rates than trypsinized single organoid cells. These were the steps we followed to express EB1ΔC-2xEGFP in organoids: (1) Add Dispase (1 mg/ml) to each well to dissolve the Matrigel at room temperature for 1 hour. (2) Spin down (at 1,000 rpm for 4 minutes) and mix organoids with 10 µL of viral particles (enough for 1 well with three Matrigel drops of 30-40 µL with organoids containing 1-2 million cells) with 10 mg/mL Y27632 ROCK inhibitor and 10 mg/mL polybrene for 30 minutes. (3) Spin the organoids with viral particles for 1 hour at 600 g. (4) Leave the organoids for a 6-hour incubation. (5) Spin down and plate in Matrigel. (6) 1 hour later, add medium. We used 0.5 mg/mL blasticidin for only one round of medium (3 days) because an increase of the density of labeled cells in the organoids reduced our ability to image MT tips with good contrast.

To optimize the drug regimens, we performed titration experiments (0.1 nM, 0.2 nM, 0.5 nM, 1 nM, 10 nM, and 100 nM) with the six tubulin inhibitors to obtain the drug concentration at which the cells from each of the breast cancer cell lines in Table 2 remain viable for over 168 hours after treatment. Docetaxel, ixabepilone, vinorelbine, vincristine, and vinblastine were obtained from Sigma-Aldrich in powder form. Paclitaxel and eribulin (in intravenous solution) were a gift from Linda Vahdat. Sorafenib and osimertinib (at 100 nM, 1 µM, and 10 µM concentrations) were a gift from Andrei Goga. BI 2536 at concentration of 250 nM was used for PLK1 inhibition.

**Microscopy Imaging**

We imaged EB1ΔC-2xEGFP-expressing organoids (Koo et al., 2013) using time-lapse spinning disk confocal microscopy. For this imagine, we used a long working distance 60x magnification water immersion objective with an NA of 1.45 and a 100x magnification oil immersion objective of an NA of 1.49 NA for all cultured cells. We acquired images every half a second for a minute to collect datasets without photobleaching (Gierke and Wittmann, 2012).

**Abbreviations**

BC – breast cancer

CLASP – cytoplasmic linker-associated protein

CLIP170 – cytoplasmic linker protein 170

DAPI – 4′,6-diamidino-2-phenylindole

EB1 – microtubule end binding protein, 1

EB1ΔC – EB1 construct truncated at amino acid 248 that does not interact with other proteins at the microtubule end

EGFP – enhanced green fluorescent protein (2xEGFP indicates two molecules)

ER – estrogen receptor

FACS – fluorescence-activated cell sorting

GFP – green fluorescent protein

GTPase – guanosine triphosphatase

HER2 – human epidermal growth factor receptor, 2

Mad2 – mitotic arrest deficient protein, 2

MAP – microtubule-associated protein

MAP2/tau – microtubule-associated protein, 2

MT – microtubule

$NAD^+$ – nicotinamide adenine dinucleotide, oxidized form

Op18 – oncoprotein 18 (stathmin)

PCa – prostate cancer

PLK1 – polo-like kinase, 1

pVHL – von Hippel-Lindau protein

ROCK – Rho-associated coiled-coil kinase

TNBC – triple-negative breast cancer

**Ethics Declaration**

Approval of tissue requests #14-04 and #16-05 to the UCSF Cancer Center Tissue Core and the Genitourinary Oncology Program was given.

**Data Availability Statement**

The datasets used and/or analyzed during the current study are available from the corresponding author upon reasonable request.

**Conflicts of Interest**

The author declares no conflicts of interest.

**Funding Statement**

No funding was received to assist with the preparation of this manuscript.

## ACKNOWLEDGEMENTS

I thank Julia Rohrberg for giving me the opportunity to collect image datasets for this manuscript while I was imaging cells for (Rohrberg et al., 2020) and for the Myc knockdown cells. I also thank Torsten Wittmann, Thomas Maresca, and Aaron Groen for discussing cytoskeleton regulation with me, and Johan de Rooij for labeling the colon cancer organoid with EB1ΔC-2xEGFP. The computational results in Table 2 were obtained by Jesús Acevedo, a summer exchange student, and Daniel Bidikov, a summer high school student, while I was training them in using ClusterTrack (Matov et al., 2010) on image datasets I obtained with the help of Benet Pera. I am grateful to the Genitourinary Tissue Utilization committee and the Genitourinary and Prostate SPORE Tissue Cores at the UCSF Cancer Center for the approval of my tissue requests #14-04 and #16-05 and the Stand Up To Cancer / Prostate Cancer Foundation (SU2C/PCF) West Coast Dream Team (WCDT).

## SUPPLEMENTARY MATERIALS

Video 1 – Time-lapse movie of a rocking mitotic spindle of an MDA-MB-231 cell with labeled EB1 comets (example #1). https://vimeo.com/382125072/bce76f2ef0

Video 2 – Time-lapse movie of a rocking mitotic spindle of an MDA-MB-231 cell with labeled EB1 comets (example #2). https://vimeo.com/382256739/ccd8ab1b63

Video 3 – Time-lapse movie of colorectal organoid with a rocking mitotic spindle with labeled EB1 comets. https://vimeo.com/995869434/e93d5ae421

Video 4 – Time-lapse movie of colorectal organoid with tracking of EB1 comets. Different colors are used for different trajectories. All trajectories are visible only during periods of MT polymerization and are removed from the screen afterwards to improve visibility. https://vimeo.com/995872753/173a2843bf